\documentclass{article}

\usepackage{arxiv}

\usepackage[utf8]{inputenc} % allow utf-8 input
\usepackage[T1]{fontenc}    % use 8-bit T1 fonts
\usepackage{url}            % simple URL typesetting
\usepackage{booktabs}       % professional-quality tables
\usepackage{amsfonts}       % blackboard math symbols
\usepackage{nicefrac}       % compact symbols for 1/2, etc.
\usepackage{microtype}      % microtypography
\usepackage{lipsum}		% Can be removed after putting your text content
\usepackage{graphicx}
\usepackage{doi}

\title{Ensemble  CNN models for Covid-19 Recognition and  Severity Perdition From 3D CT-scan}

\author{ \href{https://sciprofiles.com/profile/FaresBougourzi}{\includegraphics[scale=0.06]{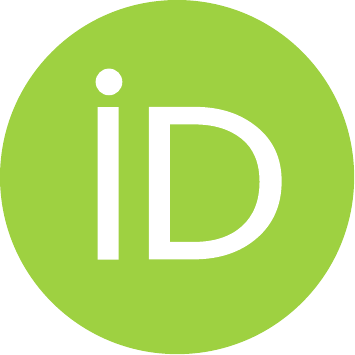}\hspace{1mm}Fares BOUGOURZI}
\thanks{.} \\
Institute of Applied Sciences and Intelligent Systems, \\
	National Research Council of Italy (CNR),\\
	  73100 Lecce, Italy \\
	\texttt{faresbougourzi@gmail.com} \\
	\And
	\href{https://orcid.org/0000-0000-0000-0000}{\includegraphics[scale=0.06]{orcid.pdf}\hspace{1mm}Cosimo Distante} \\
	Institute of Applied Sciences and Intelligent\\ Systems, National Research Council of Italy,\\ 73100 Lecce, Italy \\ Department of Innovation Engineering, University \\of Salento, 73100 Lecce, Italy \\
	\texttt{cosimo.distante@cnr.it} \\	
	\And
	\href{https://orcid.org/0000-0000-0000-0000}{\includegraphics[scale=0.06]{orcid.pdf}\hspace{1mm}Fadi DORNAIKA} \\
	University of the Basque Country UPV/EHU,\\
	San Sebastian, SPAIN; IKERBASQUE, Basque \\ Foundation for Science, Bilbao, SPAIN \\
	\texttt{fadi.dornaika@ehu.eus} \\
	\And
	\href{https://orcid.org/0000-0000-0000-0000}{\includegraphics[scale=0.06]{orcid.pdf}\hspace{1mm}Abdelmalik Taleb-Ahmed} \\
	IEMN UMR CNRS 8520, Université \\Polytechnique Hauts de France, UPHF\\
	\texttt{Abdelmalik.Taleb-Ahmed@uphf.fr} \\
}

\renewcommand{\shorttitle}{\textit{arXiv} Template}

\hypersetup{
pdftitle={A template for the arxiv style},
pdfsubject={q-bio.NC, q-bio.QM},
pdfauthor={David S.~Hippocampus, Elias D.~Striatum},
pdfkeywords={First keyword, Second keyword, More},
}

\begin{document}
\maketitle

\begin{abstract}
Since the appearance of Covid-19 in late 2019, Covid-19 has become an active research topic for the artificial intelligence (AI) community. One of the most interesting AI topics is Covid-19 analysis of medical imaging. CT-scan imaging is the most informative tool about this disease. 

This work is part of the 2nd COV19D competition, where two challenges are set: Covid-19 Detection and Covid-19 Severity Detection from the CT-scans. 
For Covid-19 detection from CT-scans, we proposed an ensemble of 2D Convolution  blocks with Densenet-161 models. Here, each  2D convolutional block with Densenet-161 architecture is trained separately and in  testing phase, the ensemble model is based on the average of their probabilities. On the other hand, we proposed an ensemble of Convolutional Layers with Inception models for Covid-19 severity detection. In addition to the Convolutional Layers, three Inception variants were used, namely Inception-v3, Inception-v4 and Inception-Resnet. 

Our proposed approaches outperformed the baseline approach in the validation data of the 2nd COV19D competition by 11\% and 16\% for Covid-19 detection and Covid-19 severity detection, respectively.

\end{abstract}

% keywords can be removed

\section{Introduction}
Since the appearance of the Covid-19 pandemic in the late of 2019, reverse transcription-polymerase chain reaction (RT-PCR) has been considered as the golden standards for Covid-19 Detection. However, the RT-PCR test has many drawbacks including inadequate supply of RT-PCR kits, time-consuming consumption, and considerable false negative results \cite{jin_rapid_2020, wu_jcs_2021, kucirka_variation_2020}. To deal with this limitations, Medical Imaging modalities have been widly used as supporting tools. These imaging modalities include X-rays and CT-scans \cite{vantaggiato2021covid,bougourzi_recognition_2021}. Indeed, CT-scans are not only used to detect Covid-19 infected cases, but they could be used to follow up the state of the patient and predicting the disease severity \cite{bougourzi_per-covid-19_2021, bougourzi_challenge_2021}.

In the last decade, Deep Learning methods have become dominant in most of the computer vision tasks and they have achieved high performance  compared to traditional methods \cite{bougourzi_fusing_2020, bougourzi_deep_2022}. However, the main drawback of deep learning, especially the CNN architecture is the need of huge labelled data, which is hard to be obtained in medical domains \cite{bougourzi_per-covid-19_2021}. On the other hand, most of the proposed CNN architectures adopted for Static images (single image input) \cite{bougourzi_challenge_2021}.

In this work, we adopted pretrained CNN architectures to detect Covid-19 infection and Covid-19 Severity from 3D CT-scans as part  of 2nd COV19D Competition. furthermore, we trained a CNN model to filter the slices that do not show lungs region. For the Covid-19 Severity detection, Att-Unet model were trained to segment the lungs regions to remove non-important features. In addition, two volume input branches were used to reduce the lost information due to the resizing transformation and the variation of the number of slices from one CT-scan to another. 
The main contributions of this paper can be summarized as follows: 

\begin{itemize} 
\item For Covid-19 detection from CT scans, we proposed an ensemble of Convolution 2D blocks with Densenet-161 models. Each convolution 2D block with densenet-161 architecture was trained separately, and the ensemble model is based on the average of their probabilities.

\item For detecting the severity of Covid-19 from CT scans, we proposed an ensemble of Convolutional Layer with Inception models. Three Inception variants were used in addition to the Convolutional layers, namely Inception-v3, Inception-v4, and Inception-Resnet. The Inception variants were trained twice separately and the ensemble model is the average of the  probabilities of the six models.

\item Our proposed approaches outperformed the baseline approach in the validation data of the 2nd COV19D competition by 11\% and 16\% for Covid-19 detection and Covid-19 severity detection, respectively.

\item The codes used and the pre-trained models are publicly available in \url{https://github.com/faresbougourzi/2nd-COV19D -Competition}. ( Last accessed on June, 28{$^{th}$} 2022).

\end{itemize}

This paper is organized in following way:  Section \ref{S:1} describes our proposed approaches for Covid-19 Detection and Severity Detection. The experiments and results are described in Section \ref{S:2}. Finally, we concluded our paper in Section \ref{S:3}.

%%%%%%%%%%%%%%%%%%%%%%%%%%%%%%%%%%%%%
\section{Our Approaches}
\label{S:1}

In the following two sections, we will describe our proposed approaches for Covid-19 Recognition and Severity prediction, respectively.

%ResneXt-50~, %Densenet-161~, %Inception-v3~\cite{Inception}, %and~Wide-Resnet-50~\cite{wideresnet} 
\subsection{Covid-19 Detection}
\label{sec:recapp}

\begin{figure}
    \includegraphics[width = 7in, height = 4in]{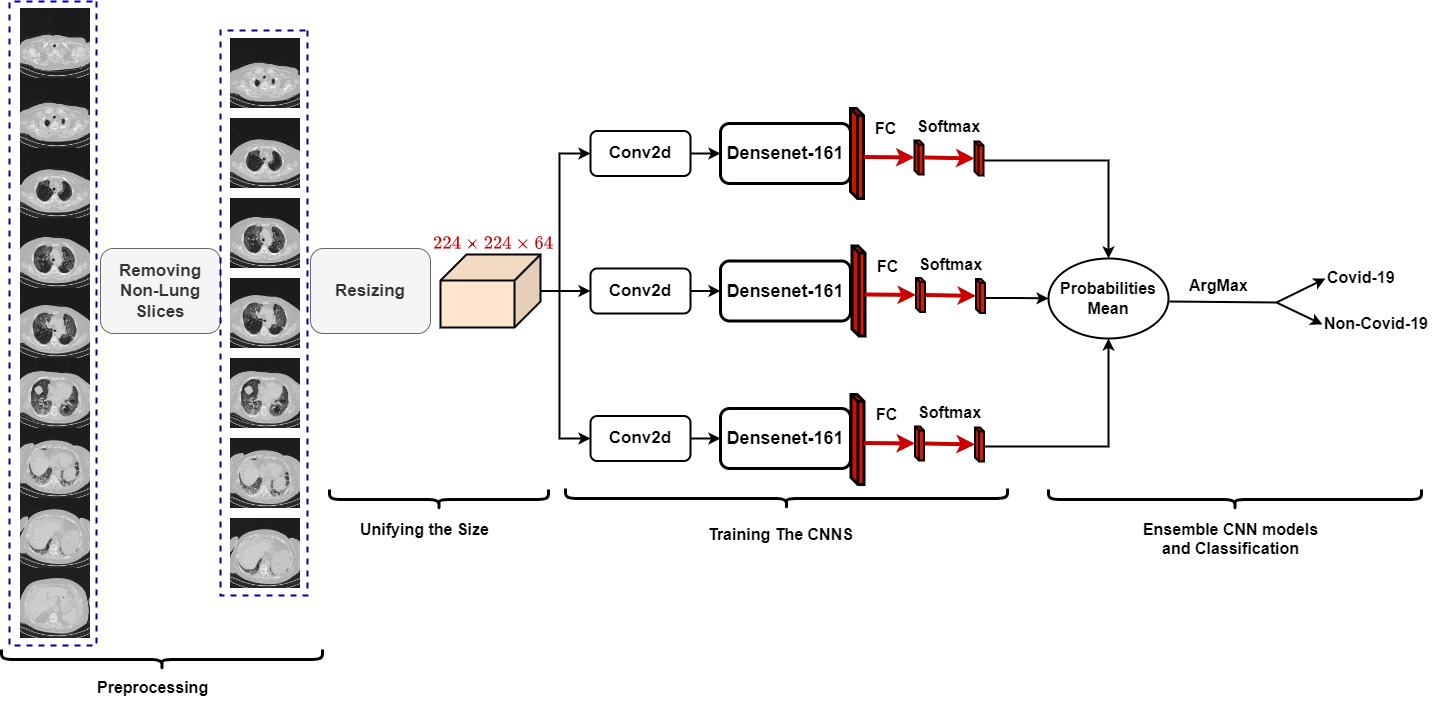} 
    \caption{Our proposed Covid-19 Detection Approach.}
    \label{fig:approach1}
\end{figure}

Our proposed approach to Covid-19 Detection for the 2nd COV19D competition is summarized in Figure \ref{fig:approach1}. Because the 3D CT-scans have a different number of slices depending on the scanner and acquisition settings and contain multiple non-lung slices, we trained a model to remove the non-lung slices. For this purpose, we manually labelled 20 CT-scans from the training data as lung and non-lung. In addition, we used the following datasets: COVID-19 CT segmentation \cite{ COVID-19-Dataset},  Segmentation dataset nr.2 \cite{COVID-19-Dataset}, and COVID-19-CT Seg dataset \cite{2021MedPh..48.1197M} to have more labelled data as lung and non-lung slices. 
After creating the lung and non-lung sets, we trained the ResneXt-50 model \cite{Resnext} to remove the slices that did not show lungs at all. Since the CT scans have different numbers of slices, we concatenated all lung slices and then reduced them to $224\times224\times64$ as shown in Figure \ref{fig:approach1}. To distinguish between Covid-19 and Non-Covid-19 CT scans, we separately trained three Densenet-161 \cite{Densenet} models. Since the input size of Denesenet-161 model is an RGB image and we have a 3D volume of size $224\times224\times64$, we added Convolution block with a 3 x 3 kernel, a stride of 1, and a padding of 1. The 3 x 3 convolution block takes 64 input channels and produces 3 output channels. It should be noted that each CNN model was trained separately. In the decision or test phase, we propose to form an ensemble of the three trained CNN architectures by averaging their probabilities.

%%%%%%%%%%%%%%%%%%%%%%%%%%%%%%%%%%%%%%%%%%%%%%%%%%%%
%%%%%%%%%%%%%%%%%%%%%%%%%%%%%%%%%%%%%%%%%%%%%%%%%%[h]

\subsection{Covid-19 Severity Detection}
\label{sec:appsev}
\begin{figure}
    \includegraphics[width = 7in, height = 4in]{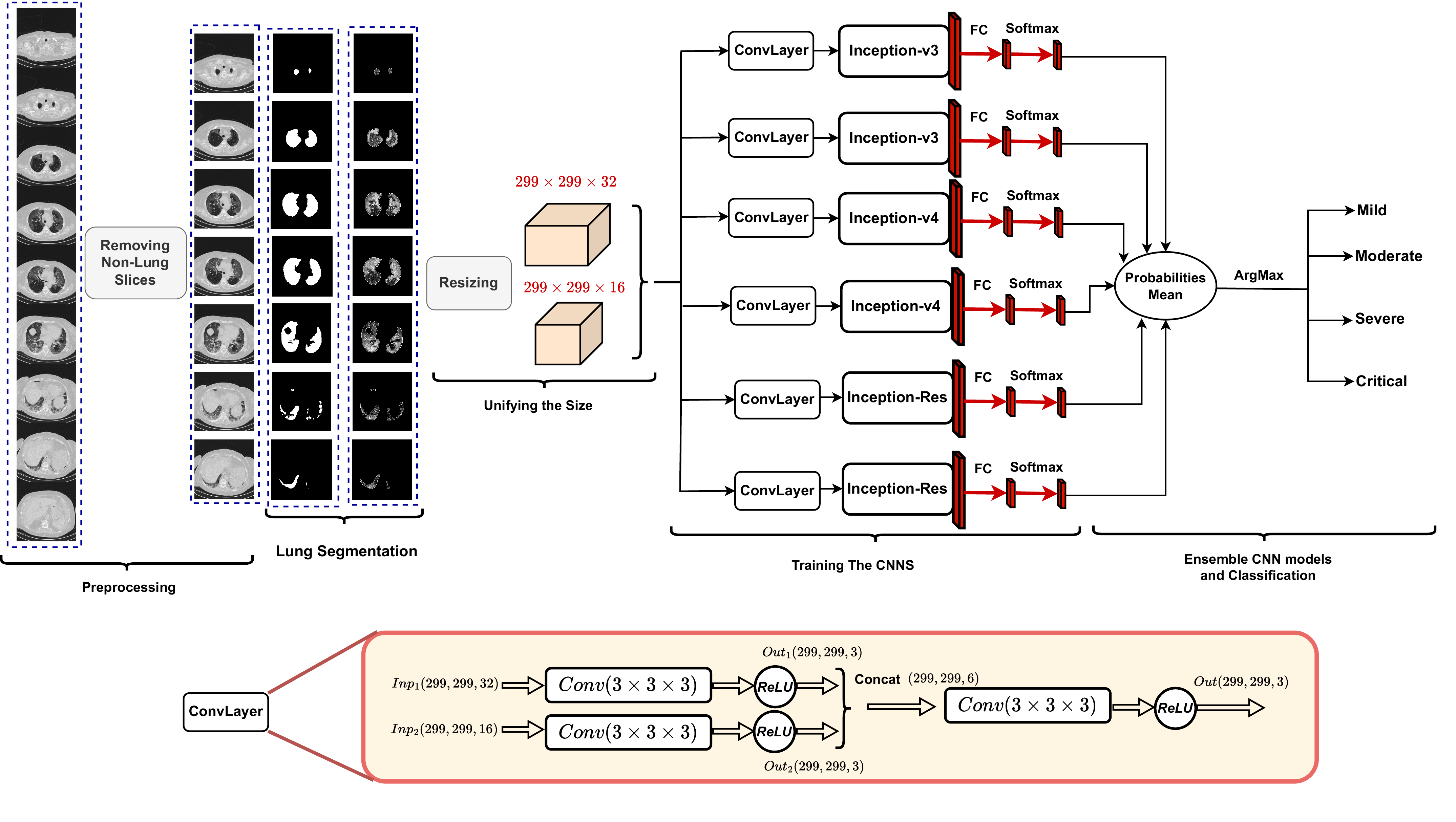} 
    \caption{Our proposed Covid-19 Severity Detection Approach.}
    \label{fig:approach2}
\end{figure}

Our proposed approach for Covid-19 severity prediction for the 2nd COV19D competition is summarized in Figure \ref{fig:approach2}. Similar to the Covid-19 detection approach in section \ref{sec:recapp}, we used the trained ResneXt-50 model to remove non-lung slices.
To segment the lungs, we trained the Att-Unet \cite{oktay_attention_2018} architecture using the following datasets COVID-19 CT segmentation \cite{ COVID-19-Dataset},  Segmentation segmentation dataset nr.2 \cite{COVID-19-Dataset}, and COVID-19-CT-Seg dataset \cite{2021MedPh..48.1197M}. After removing the non-lung slices and segmenting the lung regions to remove unnecessary features, we resized the obtained slices into two volumes $224\times224\times32$ and $224\times224\times16$. The goal of the two volumes is to reduce the information lost due to resizing and to obtain two views of each CT scan. The two input volumes were fed into the convolution block. The convolution Layer consists of three 3 by 3 convolution blocks with stride 1 and badding 1. The first convolution block transforms the volume $224\times 224\times 32$ to $224\times224\times3$. Similarly, the second convolution block transforms the volume $299\times299\times16$ into $299\times299\times3$. The third convolution block transforms the concatenation of the outputs of the first and second convolution blocks ($299\times299\times3$) into $224\times224\times3$. The output volume of the third convolution block is used as input to the architecture. In our experiments, we used Inception architectures as CNN backbones, namely inception-v3, inception-V4, and Inception-Resnet \cite{Inception}. Our approach to predict Covid-19 severity consists of six ConvLayers + CNN branches, as shown in Figure \ref{fig:approach2}, where each branch was trained separately. To create a more comprehensive ensemble model, we trained each CNN model twice. The final Covid-19 severity prediction was obtained by averaging the prediction probabilities of the six models.

%%%%%%%%%%%%%%%%%%%%%%%%%%%%%%%%%%%%%%%%%%%%%%%%%%%%
%%%%%%%%%%%%%%%%%%%%%%%%%%%%%%%%%%%%%%%%%%%%%%%%%%[h]
%%%%%%%%%%%%%%%%%%%%%%%%%%%%%%%%%%%%%%%%%
\section{Experiments and Results}
\label{S:2}
\subsection{The COV19-CT-DB Database}
The COVID19-CT-Database (COV19-CT-DB) \cite{kollias2022ai,kollias2021mia,kollias2020deep, kollias2020transparent, kollias2018deep}  consists of chest CT scans that are annotated for the existence of COVID-19.

%%%%%%%%%%%%%%%%%%%%%%%%%%%%%%%%%%%%%%%%%%%%
\begin{table}[h]
 \caption{Data samples in each Set. }
\label{tab:datadet}
\centering
\begin{tabular}{|p{3cm}|p{2cm}|p{2cm}|}
\hline

\textbf{Set}  & \textbf{Training} &\textbf{Validation}  \\\hline
Covid-19& 882 & 215  \\ \hline 
Non-Covid-19& 1110 &289  \\ \hline  

\end{tabular}
\end{table}
%%%%%%%%%%%%%%%%%%%%%%%%%%%%%%%%%%%%%%%%%%%%

%%%%%%%%%%%%%%%%%%%%%%%%%%%%%%%%%%%%%%%%%%%%
\begin{table}[h]
 \caption{Data samples in each Severity Class }
\label{tab:datasev}
\centering
\begin{tabular}{|p{3cm}|p{2cm}|p{2cm}|}
\hline

\textbf{Severity Class}  & \textbf{Training} &\textbf{Validation}  \\\hline
1& 85 &  22 \\ \hline 
2& 62& 10\\ \hline
3& 85& 22  \\ \hline 
4& 26& 5 \\ \hline 

\end{tabular}
\end{table}
%%%%%%%%%%%%%%%%%%%%%%%%%%%%%%%%%%%%%%%%%%%%
  
COVID19-CT-DB contains 3-D CT scans of the chest, which were analysed for the presence of COVID-19. Each of the 3-D scans contains a different number of slices, ranging from 50 to 700. 
The database was divided into a training set, a validation set, and a test set. The training set contains a total of 1992 3-D CT scans. The validation set consists of 494 3-D CT scans. The number of COVID-19 and non COVID-19 cases in each set is given in Table \ref{tab:datadet}. 

Further partitioning of the COVID-19 cases was based on the severity of COVID-19 , which was given by four medical experts in a range of 1 to 4, with 4 denoting critical status. 
The training set contains a total of 258 3-D CT scans. The validation set consists of 61 3-D CT scans. The number of scans in each severity class  in these sets is shown in Table \ref{tab:datasev}.

\subsection{Experimental Setup}
For Deep Learning training and testing, we used the Pytorch \cite{paszke_pytorch_2019} Library with NVIDIA GPU Device GeForce TITAN RTX 24 GB. The batch size used consists of 16 CT-scan volumes for both tasks. We trained the networks for 40 epochs. The initial learning rate is 0.0001, which decreases by 0.1 after 15 epochs, followed by another 0.1 decrease after 30 epochs.

%%%%%%%%%%%%%%%%%%%%%%%%%%%%%%%%%%%%%%%%%%%%
%%%%%%%%%%%%%%%%%%%%%%%%%%%%%%%%%%%%%%%%%%%%
\subsection{Covid-19 Recognition}
\label{sec:headings}

Table \ref{tab:Recresult} summarizes the Covid-19 detection results (F1 score) obtained on the validation data.
As described in section \ref{sec:recapp}, we trained three Convolutional Block + Densenet-161 architectures. For simplicity, we refer to Convolutional Block + Densenet-161 with Densenet-161 and the training iteration. 
Comparing with the baseline results, we find that all the trained Densenet models outperform the baseline approach by about 11\%, 9\% and 10\% for model 1, 2 and 3, respectively. Moreover, the ensemble of these Densenet models improves the result compared to the best trained model (Model 1).

%%%%%%%%%%%%%%%%%%%%%%%%%%%%%%%%%%%%%%%%%%%%
\begin{table}[h]
 \caption{Covid-19 Detection Results. }
\label{tab:Recresult}
\centering
\begin{tabular}{|p{1cm}|p{4cm}|c|}
\hline

\textbf{Model}  &\textbf{Architecture}  & \textbf{Macro F1-Score}   \\\hline
- &Baseline& 77  \\ \hline 
1&Densenet-161 Model1& 88.34  $\pm$ 0.59\\ \hline
2&Densenet-161 Model2& 86.91  $\pm$ 0.59\\ \hline 
3&Densenet-161 Model3& 87.46   $\pm$ 0.59\\ \hline 
4&Ensemble Models&  88.84  $\pm$ 0.31\\ \hline 
\end{tabular}
\end{table}

%%%%%%%%%%%%%%%%%%%%%%%%%%%%%%%%%%%%%%%%%%%%

\subsection{Covid-19 Severity Detection}
\label{sec:headings}

Table \ref{tab:Sevresult} summarizes the results (F1 score) of detecting the severity of Covid-19 using the validation data. As described in section\ref{sec:appsev}, we trained six convolutional layers + CNN backbones. The CNN backbones used are Inception-v3, Inception-V4, and Inception-Resnet, and each model was trained twice to obtain a greater variety of predictors. For simplicity, we refer to ConvLayer + CNN backbone with the backbone architecture name. 
Compared to the baseline results, we find that our trained models achieve better performance of 11\%, 11\%, 12\%, 10\%, 10\% and 9\% for model 1-6, respectively. Moreover, the ensemble of trained models improves the result compared to all models and achieves 16\% better result than the baseline approach.

%%%%%%%%%%%%%%%%%%%%%%%%%%%%%%%%%%%%%%%%%%%%
\begin{table}[h]
 \caption{Covid-19 Severity Detection Results. }
\label{tab:Sevresult}
\centering
\begin{tabular}{|p{1cm}|p{4cm}|c|}
\hline

\textbf{Model}  &\textbf{Architecture}  & \textbf{Macro F1-Score}   \\\hline
-&Baseline& 63  \\ \hline 
1&Inception-v3 Model1& 74.86 $\pm$ 0.5\\ \hline
2&Inception-v3 Model2& 74.75 $\pm$ 0.5 \\ \hline 
3&Inception-v4 Model1& 75.60 $\pm$ 1.1\\ \hline
4&Inception-v4 Model2& 73.48 $\pm$ 1.1\\ \hline
5&Inception-Res Model1& 73.37 $\pm$ 0.7\\ \hline 
6&Inception-Res Model2& 72.01 $\pm$ 0.7\\ \hline
7&Ensemble Models& 79.10 $\pm$  0.6\\ \hline 
\end{tabular}
\end{table}

%%%%%%%%%%%%%%%%%%%%%%%%%%%%%%%%%%%%%%%%%%%%

%%%%%%%%%%%%%%%%%%%%%%%%%%%%%%%%%%%%%%%%%%%%
\section{Conclusion}
\label{S:3}
In this work, we proposed two ensemble CNN-based approaches for the 2nd COV19D competition. 
For Covid-19 detection from CT scans, we proposed an ensemble of 2D convolution blocks with Densenet-161 models. Here, the individual convolution 2D blocks with Densenet-161 architecture are trained separately, and the ensemble model is based on the average of their probabilities at testing phase. On the other hand, we proposed an ensemble of Convolutional Layers with Inception models for Covid-19 severity detection. In addition to the Convolutional Layers, three Inception variants were used, namely Inception-v3, Inception-v4, and Inception-Resnet. 

Our proposed approaches outperformed the baseline approach by a significant margin in the validation data of the 2nd COV19D competition with 11\% and 16\% in Covid-19 detection and Covid-19 severity detection, respectively. In future work, we will evaluate our approaches in other datasets and use the attention mechanism to select the representative slices for Covid-19 detection.

%\bibliographystyle{unsrt}
%\bibliography{references}

\end{document}